\documentclass[prl,aps,twocolumn,showpacs,floatfix]{revtex4}
\usepackage{graphicx}

\begin{document}
\title{Non-monotonicity in the quantum-classical transition: \\
Chaos induced by quantum effects} 
\author{Arie Kapulkin$^{(a)}$ and Arjendu K. Pattanayak$^{(b)}$} 
\date{June 2007}
\affiliation{(a) 128 Rockwood Cr, Thornhill, Ont L4J 7W1 Canada
\\
(b) Department of Physics and Astronomy, Carleton College, Northfield, 
Minnesota 55057} 
\begin{abstract}
The transition from classical to quantum behavior for chaotic systems 
is understood to be accompanied by the suppression of chaotic effects 
as the relative size of $\hbar$ is increased. We show evidence to the 
contrary in the behavior of the quantum trajectory dynamics of a 
dissipative quantum chaotic system, the double-well Duffing oscillator. 
The classical limit in the case considered has regular behavior, but 
as the effective $\hbar$ is increased we see chaotic behavior. This chaos 
then disappears deeper into the quantum regime, which means that the 
quantum-classical transition in this case is non-monotonic in $\hbar$.
\end{abstract}
\maketitle

Open nonlinear quantum systems are critical in understanding the 
foundations of quantum behavior, particularly the transition 
from quantum to classical mechanics. For example, density matrix
formulations have been used to argue that quantum systems decohere 
rapidly when the classical counterpart is chaotic, with the 
decoherence rate determined by the classical Lyapunov exponents of 
the system\cite{zp}. This applies to entanglement and fidelity 
issues as well\cite{jacquod,jalabert, arul}, since decoherence 
amounts to entanglement with the environment.

A powerful alternative way of studying open quantum systems is the Quantum 
State Diffusion (QSD) approach\cite{percival}. This approach enabled the 
resolution of an important paradox, namely that in the absence of a QSD-like 
formulation, classical chaos cannot be recovered from quantum mechanics, 
indicating that the $\hbar \to 0$ limit is singular. Early QSD 
work\cite{brun} studied the convergence towards classical trajectories for 
a chaotic system, considering quantum Poincar\'e sections of the 
quantities $\langle \hat x \rangle$ and $\langle \hat p \rangle$. 
It showed that the classical chaotic attractor is recovered when the 
system parameters were such that $\hbar$ was small relative to the system's 
characteristic action. As the relative $\hbar$ increased, the attractor 
disappeared gradually, suggesting a persistence of chaos into the quantum 
region, consistent with later, more quantitative analyses~\cite{ota,Adamyan}. 
Related work\cite{salman} studied a quantum system that is being 
continuously weakly measured, which leads to similar equations as those 
for QSD\cite{Steck}. This also showed that chaos is recovered in the 
classical limit, and that it persists, albeit reduced, substantially into 
the quantum regime. Another related study~\cite{everitt} of coupled 
Duffing oscillators, showed that quantum effects, specifically entanglement, 
persist in a quantum system even when the system is classical enough to
be chaotic.

While the quantum persistence of chaos is interesting, it is still 
consistent with the understanding that chaos is a classical phenomenon that 
is suppressed quantum mechanically. Do quantum effects always decrease chaos, 
however? A closed Hamiltonian quantum system studied within a gaussian 
wavepacket approximation~\cite{akp-schieve} manifested chaos absent in its 
classical version. This has been understood to be an artifact of the 
approximation, since the full quantum system is not chaotic. Follow-up 
work with an open system~\cite{liu-schieve} also manifested quantum chaos, 
but it is not clear if this was not due to the approximations made. 

In this paper we show the first (to the best of our knowledge) evidence 
of chaos being {\bf induced} by quantum effects using QSD, whence there
are no artifacts of approximations. Specifically, in a system with a 
non-chaotic classical limit, as we increase the relative $\hbar$, chaos 
emerges, due to explicitly quantum efffects (tunneling and zero-point 
energy) and as $\hbar$ is increased further, the chaos disappears. This 
intriguing result is arguably relatively common. More broadly, it shows 
that the quantum-classical transition for nonlinear systems is {\bf not} a 
monotonic function of $\hbar$. 
\begin{figure}[htbp]
\centerline{\includegraphics[width=8.3cm,height=10.8cm,clip]
{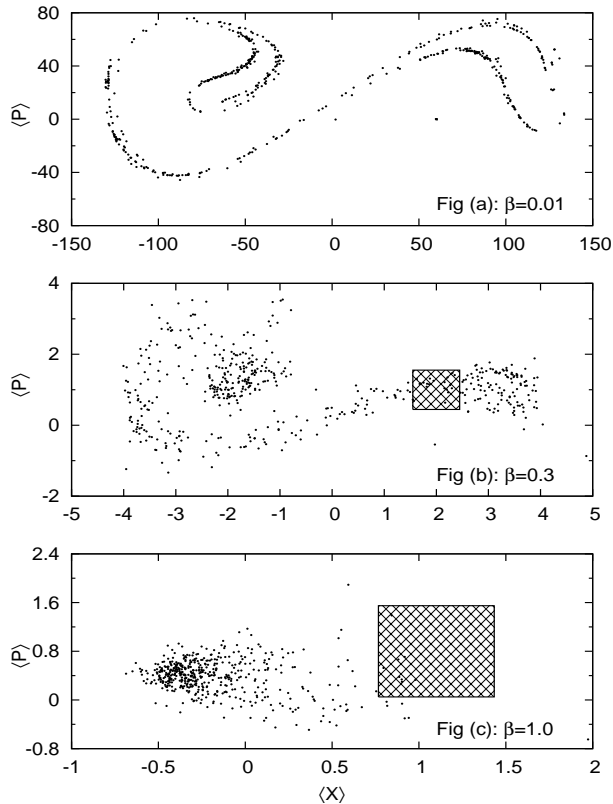}}
\caption{Poincar\'e sections for $\Gamma=0.125$, and $\beta=0.01,0.3,1.0$ 
reading from top to bottom. The monotonic transition from classical chaos 
to quantum regularity is to be contrasted with the non-monotonicity in 
Fig.~(3). The shaded squares (area equal to $1$) in the center right of
each graph indicate the Planck cell of size $\hbar$.} 
\label{figone}
\end{figure}
In QSD, the evolution equation for a realization $|\psi\rangle$ of the 
system interacting with a Markovian environment is 
\begin{eqnarray}
|d\psi\rangle &=& -\frac{i}{\hbar}\hat H |\psi\rangle dt \nonumber\\
&+& \sum_j \bigg (
\langle \hat L_j^\dagger \rangle \hat L_j
-\frac{1}{2}\hat L_j^\dagger \hat L_j
-\frac{1}{2}\langle \hat L_j^\dagger \rangle \langle \hat L_j\rangle
\bigg )
|\psi\rangle dt\nonumber\\
&+&\sum_j (\hat L_j - \langle L_j \rangle )|\psi \rangle d\xi_j
\label{sse}
\end{eqnarray}
where the Lindblad operators $\hat L_j$ model coupling to an external 
environment. The density matrix is recovered as the ensemble mean $M$ 
over different realizations as $\hat \rho = M |\psi\rangle\langle \psi
|$\cite{percival}. 
The $d\xi_j$ are independent normalized complex differential random variables 
satisfying $M(d\xi_j) =0; M (d\xi_j d\xi_{j'}) =0; M (d\xi_jd\xi_{j'})
=\delta_{jj'}dt$. 

Now consider specifically the classical driven dissipative Duffing oscillator
\begin{equation}
\ddot{x} + 2\Gamma \dot{x} + x^{3} - x = g \cos(\Omega t),
\label{Eq:cDuff}
\end{equation}
a particle of unit mass in a double-well potential, with dissipation 
$\Gamma$ dissipation, and a sinusoidal driving amplitude $g$. The dynamics 
depend on the parameters $\Gamma$, $g$ and $\Omega$, with chaotic behavior 
obtaining for certain parameter ranges. Chaos is found through Poincar\'e 
maps (obtained by recording $(x,\,p)$ at time intervals of $2\pi /\Omega$) 
showing a strange attractor, or the behavior of the time-series $x(t)$, or 
through a positive Lyapunov exponent. To quantize this 
problem\cite{brun,ota}, we choose the Hamiltonian $\hat{H}$ and the Lindblad 
operator $\hat{L}$ for Eq.~(\ref{sse}) as 
$\hat{H}=\hat{H}_{D}+\hat{H}_{R}+\hat{H}_{ex}$, 
\begin{eqnarray}
\hat{H}_{D} &=& \frac{1}{2m} \hat{p}^{2} + \frac{m \omega^{2}_{0}}{4l^{2}} 
\hat{x}^{4} - \frac{m \omega^{2}_{0}}{2} \hat{x}^{2},\\
\label{Eq:qDuffH}
\hat{H}_{R}&=&\frac{\gamma}{2} (\hat{x} \hat{p} + \hat{p} \hat{x}),\\
\label{Eq:reNH}
\hat{H}_{ex}&=&- gml \omega_{0}^{2} \hat{x} \cos(\omega t),\\
\label{Eq:exH}
\hat{L}&=&\sqrt{\frac{m\omega_{0} \gamma}{\hbar}}\; \hat{x} 
+ \sqrt{\frac{\gamma}{m\omega_0\hbar}}\hat{p} 
\label{Eq:Lop}
\end{eqnarray}
After some redefinitions this reduces to the dimensionless Hamltonian 
$\hat{H}_{\beta}$ and Lindblad operator $\hat{K}$
given by $\hat{H}_{\beta} = \hat{H}_{D} + \hat{H}_{R} + \hat{H}_{ex}$ 
where $\hat{H}_{D} = \frac{1}{2} \hat{P}^{2}+\frac{\beta^{2}}{4}
\hat{Q}^{4}-\frac{1}{2} \hat{Q}^{2},
\hat{H}_{R} = \frac{\Gamma}{2} \left(\hat{Q} 
\hat{P} + \hat{P} \hat{Q} \right),
\hat{H}_{ex}=- \frac{g}{\beta} \hat{Q} \cos(\Omega t),
\hat{K} = \sqrt{\Gamma} \left( \hat{Q} + i\hat{P} \right),$
and $\Omega \equiv \omega/\omega_{0}$, $\Gamma \equiv \gamma/\omega_{0}$.
The quantity $\beta^{2} = \frac{\hbar}{ml^{2}\omega_{0}}$ determines the 
relative system size, and the degree to which quantum effects influence the 
motion. Specifically\cite{ota} the limit $\beta \to 0$ yields the classical 
Eq~(\ref{Eq:cDuff}) while increasing $\beta$ increases quantum corrections 
resulting in qualitatively different dynamics. 
\begin{figure}[htbp]
\centerline{\includegraphics[width=7.5cm,height=9.7cm,angle=-90,clip]
{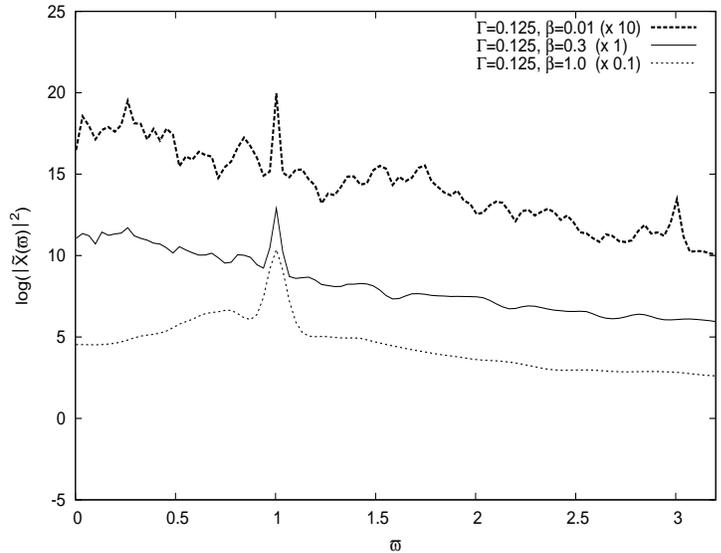}}
\caption{Low-frequency power spectra for the 3 $\beta$ values shown in 
Fig.~(1), offset for visual clarity. We see the characteristic 
rise at low frequencies\cite{everitt2} for the chaotic cases, as well 
as the monotonicity of the transition with $\beta$.}
\label{figtwo}
\end{figure}

\begin{figure}[htbp]
\centerline{\includegraphics[width=8.3cm,height=10.8cm,clip]
{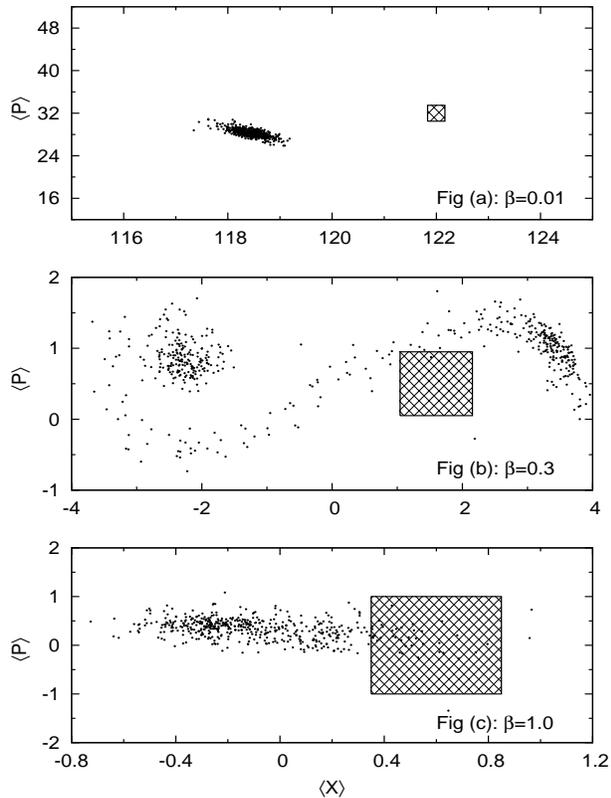}}
\caption{Poincar\'e sections for $\Gamma=0.3$, and $\beta=0.01,0.3,1.0$ 
reading from top to bottom.  The non-monotonicity of \{classical regularity 
$\to$ chaos $\to$ regularity\} is to be contrasted with the monotonicity in 
Fig.~(1).} 
\label{figthree}
\end{figure}
\begin{figure}[htbp]
\centerline{\includegraphics[width=7.5cm,height=9.7cm,angle=-90,clip]
{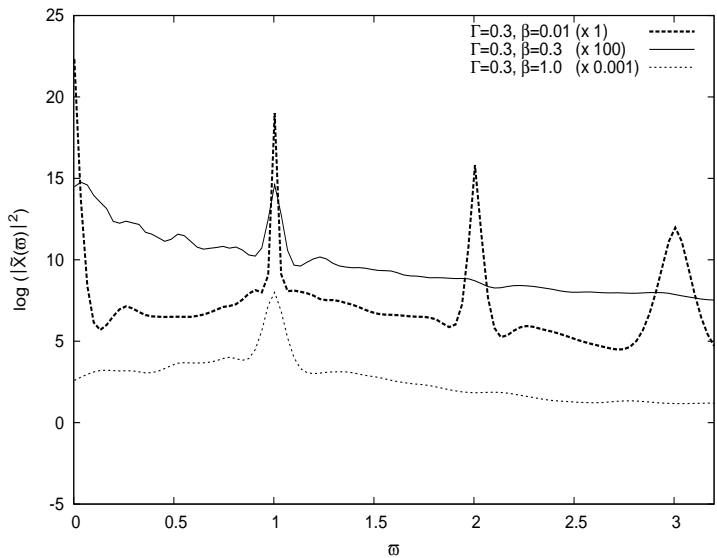}}
\caption{Low-frequency power spectra for the 3 $\beta$ values shown in 
Fig.~(3), offset for visual clarity. We see the characteristic rise at
low frequencies for the chaotic case, as well as the non-monotonicity 
of the transition with $\beta$, to be contrasted with the monotonicity in 
Fig.~(2).}
\label{figfour}
\end{figure}
\begin{figure}[htbp]
\centerline{\includegraphics[width=8.3cm,height=10.8cm,clip]
{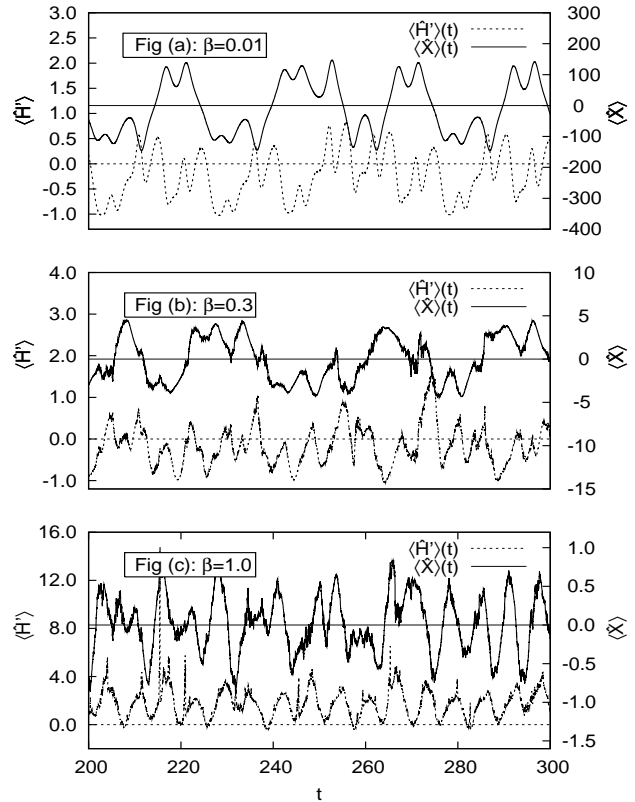}}
\caption{Expectation values of energy and position for $\Gamma=0.125$, and
$\beta=0.01,0.3,1.0$ reading from top to bottom, showing evidence for
tunneling and the effect of zero-point energy as $\beta$ increases.}
\label{figfive}
\end{figure}
\begin{figure}[htbp]
\centerline{\includegraphics[width=8.3cm,height=10.8cm,clip]
{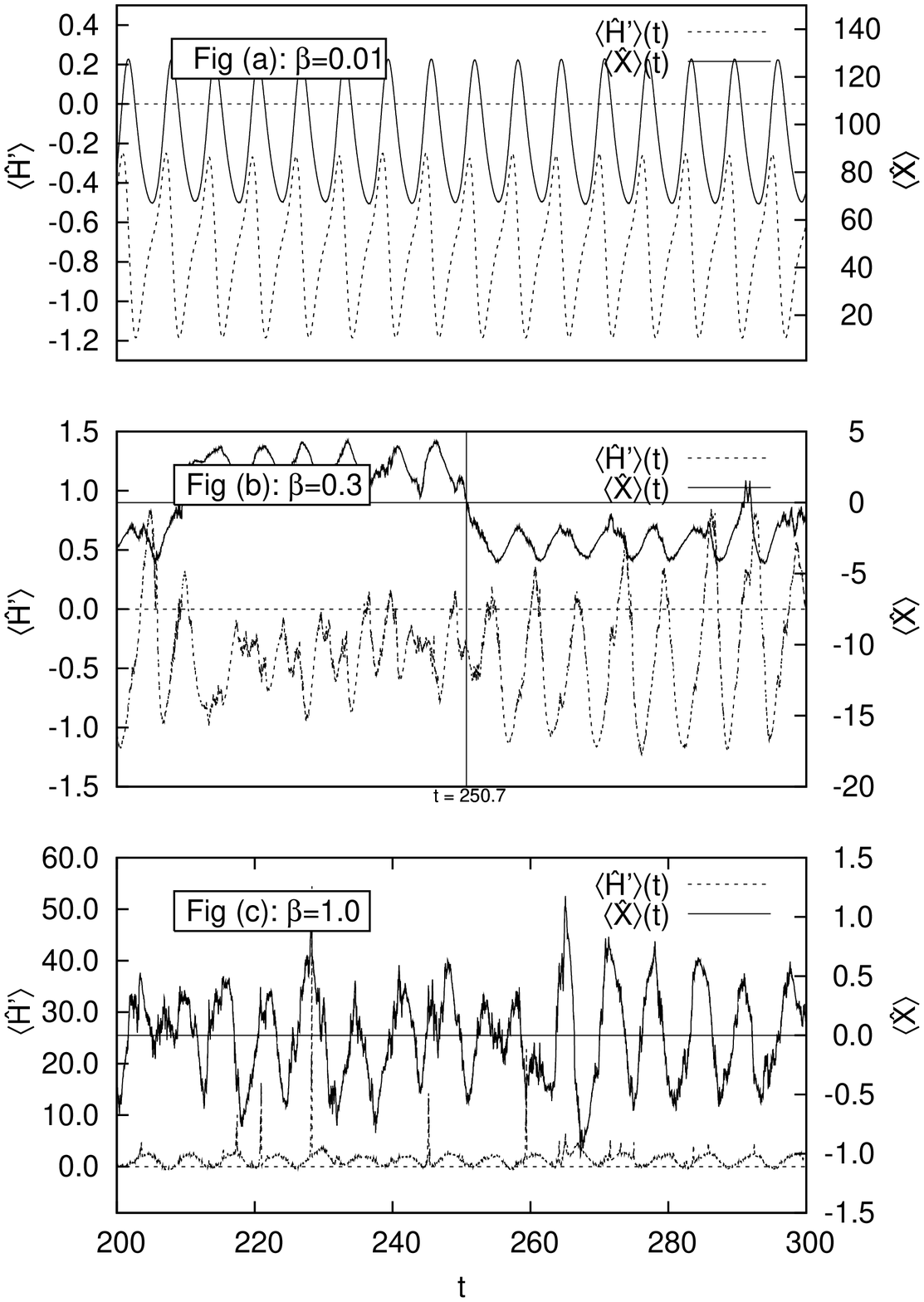}}
\caption{As in Fig.\ref{figfive} except that $\Gamma=0.3$, showing clearly 
that tunneling and the zero-point energy lead to bistability and chaos 
in this case.}
\label{figsix}
\end{figure}

We studied the quantum Duffing oscillator with simulations using the 
numerical QSD library\cite{QSD}. Previous studies\cite{brun,ota} changed 
$\beta$, with the parameters $\Gamma = 0.125$, $g=0.3$ and $\Omega=1.00$
held fixed. All these studies have the same classical limit, known 
to be chaotic\cite{Guckenheimer}. We have studied $9$ different families 
of quantum systems, each with a different classical limit, and 
examined $13$ different values of $\beta$ from essentially classical 
($\beta=0.01$) to the deep quantum regime ($\beta=1.0$). Of all these, 
we show six cases in the accompanying figures. Each simulation 
had the same initial state $|\psi(t=0)\rangle$ --- the coherent state 
$|\sqrt{2}(\langle\hat{Q}\rangle +i \langle\hat{P}\rangle)=(1.4-0.4i)\rangle$
--- and ran slightly over $500$ periods of the external driving, yielding
$500$ post-transient points for the quantum Poincar\'e sections, shown 
for various $\beta$ in Fig.~(1a,b,c) as in\cite{brun,ota}. 
We next consider the low-frequency
power spectra, shown in Fig.~(\ref{figtwo}), from the Fourier transforms 
$\tilde X (\varpi)$ of the time-series $\langle \hat Q
(t)\rangle$; we show plots smoothed via a cubic spline in Fourier space 
to focus attention on the overall trend. 

We note both the broadband contributions of the noise in Eq.~(\ref{sse}), 
and also an exponential increase in the power distribution in the 
low frequency ($\varpi\ll\Omega$) limit. This low frequency increase 
is characteristic of chaotic dynamics, and is absent in regular 
motion~\cite{everitt2}. Now consider Figs.~(3a,b,c), where we also show 
Poincar\'e sections, this time for $\Gamma = 0.3$, $g=0.3$ and $\Omega=1.00$. 
In marked contrast to the $\Gamma=0.125$ situation, in this case we see
the transition \{regular $\to$ chaos $\to$ regular\} as $\beta$ is increased. 
That is, the system becomes chaotic as quantal effects are 
increased, and then becomes regular again in the deep quantal region, 
as also evident in the power spectra in Fig.~(4). 

To understand these results, note that chaos occurs in this sort of 
system when trajectories sample the region near the unstable-fixed
point, and particularly the separatrix region, leading to intra-well 
transitions. Classically, this means that a minimum amount of energy is
needed for chaos to occur. The difference between (1a) and (3a) is due 
to the larger dissipation in (3a), confining the system to one well. The 
external driving force in the latter case is insufficient to overcome 
the potential barrier, and without this, chaos does not occur. 
Increasing $\beta$, however, increases the size of the wave-packet 
moving in the classical potential, adding extra degrees of freedom (along 
with the `classical' centroid variables, there are now wave-packet 
variances) and changing how we understand the dynamics. It is in 
particular useful 
to build intuition by thinking of quantum mechanics as classical 
behavior in an effective potential. Increasing $\beta$ amounts to 
increasing quantum effects in the system effective potential\cite{reale}. 
These quantum corrections raise the effective bottom of the well through 
added zero-point energy and also modify the well-barrier, providing 
another route between the wells (tunneling). 

These effects are apparent in time-slices of the expectation values of 
the energy operator, as well as position, in Figs.~(\ref{figfive},
\ref{figsix}). In Figs~(5a,6a) we see classical behavior: transitions 
between wells only occur with positive energy. In Fig.~(6a), the 
energy is always negative, confining the system in one well. In 
Figs.~(5b,6b) quantum effects are significant and the barrier is 
softened as the zero-point energy becomes significant and the 
potential barrier at $x=0$ decreases. Now, transitions between 
wells occur even for negative energies, which we have indicated for 
(6b) with a line at $t=250.7$: this is quantum tunneling. Fig (6b) shows 
the core reason for the non-monotonic behaviour: classically forbidden 
inter-well transitions become possible due to quantum effects, allowing for 
chaos. Figs (5c,6c) show that in the deep quantum regime, 
the quantum effects are so large that the system effectively sees a single 
well potential leading to regular motion. Classically, changing $\beta$ is 
no more than changing the units of measurement for the system, or 
equivalently rescaling the system by a constant factor. Quantum dynamics 
however are sensitive to the absolute size of the system in units of 
$\hbar$. This scale dependence, rather than a variation of the dynamical 
parameters of the system as in classical chaos is what leads to the 
bifurcation signalling the onset of chaotic behaviour. 

We have therefore seen that it is possible for the dynamics to be
`quantum' (as evidenced by tunneling effects) AND 'chaotic' simultaneously,
and specifically that the quantum effects in fact induce
chaos. More broadly, quantum effects can be non-monotonic.
It is likely that this is a generic property of nonlinear systems 
described by Hilbert space trajectories. 
{\em Acknowledgements:}
AKP gratefully acknowledges the `SIT, Wallin, and Class of 1949'
sabbatical leave fellowships from Carleton.

\end{document}